\documentclass{aa-mod}
\usepackage{graphicx}
\usepackage{longtable}
\usepackage{txfonts} 
\usepackage{natbib}
\bibpunct{(}{)}{;}{a}{}{,} 

\begin{document}

\title{A possible association of the new VHE $\gamma$-ray source
HESS\,J1825--137 with the pulsar wind nebula G\,18.0--0.7}

\author{F.A. Aharonian\inst{1}
 \and A.G.~Akhperjanian \inst{2}
 \and A.R.~Bazer-Bachi \inst{3}
 \and M.~Beilicke \inst{4}
 \and W.~Benbow \inst{1}
 \and D.~Berge \inst{1}
 \and K.~Bernl\"ohr \inst{1,5}
 \and C.~Boisson \inst{6}
 \and O.~Bolz \inst{1}
 \and V.~Borrel \inst{3}
 \and I.~Braun \inst{1}
 \and F.~Breitling \inst{5}
 \and A.M.~Brown \inst{7}
 \and P.M.~Chadwick \inst{7}
 \and L.-M.~Chounet \inst{8}
 \and R.~Cornils \inst{4}
 \and L.~Costamante \inst{1,20}
 \and B.~Degrange \inst{8}
 \and H.J.~Dickinson \inst{7}
 \and A.~Djannati-Ata\"i \inst{9}
 \and L.O'C.~Drury \inst{10}
 \and G.~Dubus \inst{8}
 \and D.~Emmanoulopoulos \inst{11}
 \and P.~Espigat \inst{9}
 \and F.~Feinstein \inst{12}
 \and G.~Fontaine \inst{8}
 \and Y.~Fuchs \inst{13}
 \and S.~Funk \inst{1}
 \and Y.A.~Gallant \inst{12}
 \and B.~Giebels \inst{8}
 \and S.~Gillessen \inst{1}
 \and J.F.~Glicenstein \inst{14}
 \and P.~Goret \inst{14}
 \and C.~Hadjichristidis \inst{7}
 \and M.~Hauser \inst{11}
 \and G.~Heinzelmann \inst{4}
 \and G.~Henri \inst{13}
 \and G.~Hermann \inst{1}
 \and J.A.~Hinton \inst{1}
 \and W.~Hofmann \inst{1}
 \and M.~Holleran \inst{15}
 \and D.~Horns \inst{1}
 \and A.~Jacholkowska \inst{12}
 \and O.C.~de~Jager \inst{15}
 \and B.~Kh\'elifi \inst{1}
 \and Nu.~Komin \inst{5}
 \and A.~Konopelko \inst{1,5}
 \and I.J.~Latham \inst{7}
 \and R.~Le Gallou \inst{7}
 \and A.~Lemi\`ere \inst{9}
 \and M.~Lemoine-Goumard \inst{8}
 \and N.~Leroy \inst{8}
 \and T.~Lohse \inst{5}
 \and J.M.~Martin \inst{6}
 \and O.~Martineau-Huynh \inst{16}
 \and A.~Marcowith \inst{3}
 \and C.~Masterson \inst{1,20}
 \and T.J.L.~McComb \inst{7}
 \and M.~de~Naurois \inst{16}
 \and S.J.~Nolan \inst{7}
 \and A.~Noutsos \inst{7}
 \and K.J.~Orford \inst{7}
 \and J.L.~Osborne \inst{7}
 \and M.~Ouchrif \inst{16,20}
 \and M.~Panter \inst{1}
 \and G.~Pelletier \inst{13}
 \and S.~Pita \inst{9}
 \and G.~P\"uhlhofer \inst{1,11}
 \and M.~Punch \inst{9}
 \and B.C.~Raubenheimer \inst{15}
 \and M.~Raue \inst{4}
 \and J.~Raux \inst{16}
 \and S.M.~Rayner \inst{7}
 \and A.~Reimer \inst{17}
 \and O.~Reimer \inst{17}
 \and J.~Ripken \inst{4}
 \and L.~Rob \inst{18}
 \and L.~Rolland \inst{16}
 \and G.~Rowell \inst{1}
 \and V.~Sahakian \inst{2}
 \and L.~Saug\'e \inst{13}
 \and S.~Schlenker \inst{5}
 \and R.~Schlickeiser \inst{17}
 \and C.~Schuster \inst{17}
 \and U.~Schwanke \inst{5}
 \and M.~Siewert \inst{17}
 \and H.~Sol \inst{6}
 \and D.~Spangler \inst{7}
 \and R.~Steenkamp \inst{19}
 \and C.~Stegmann \inst{5}
 \and J.-P.~Tavernet \inst{16}
 \and R.~Terrier \inst{9}
 \and C.G.~Th\'eoret \inst{9}
 \and M.~Tluczykont \inst{8,20}
 \and G.~Vasileiadis \inst{12}
 \and C.~Venter \inst{15}
 \and P.~Vincent \inst{16}
 \and H.J.~V\"olk \inst{1}
 \and S.J.~Wagner \inst{11}}

\offprints{O.C. de Jager \email{fskocdj@puk.ac.za}}

\institute{ 
Max-Planck-Institut f\"ur Kernphysik, Heidelberg, Germany 
\and Yerevan Physics Institute, Yerevan, Armenia 
\and Centre d'Etude Spatiale des Rayonnements, CNRS/UPS, Toulouse,
France 
\and Universit\"at Hamburg, Institut f\"ur Experimentalphysik,
Hamburg, Germany
\and Institut f\"ur Physik, Humboldt-Universit\"at zu Berlin, Germany 
\and LUTH, UMR 8102 du CNRS, Observatoire de Paris, Section de
Meudon, France
\and University of Durham, Department of Physics, Durham DH1 3LE, U.K.
\and Laboratoire Leprince-Ringuet, IN2P3/CNRS, Ecole Polytechnique,
Palaiseau, France 
\and APC, Paris Cedex 05, France \thanks{UMR 7164 (CNRS, Observatoire
  de Paris)} 
\and Dublin Institute for Advanced Studies, Dublin, Ireland
\and Landessternwarte, K\"onigstuhl, Heidelberg, Germany
\and Laboratoire de Physique Th\'eorique et Astroparticules,
Universit\'e Montpellier II, France
\and Laboratoire d'Astrophysique de
Grenoble, INSU/CNRS, Universit\'e Joseph Fourier, France
\and DAPNIA/DSM/CEA, CE Saclay, Gif-sur-Yvette, France 
\and Unit for Space Physics, North-West University, Potchefstroom,
South Africa 
\and Laboratoire de Physique Nucl\'eaire et de Hautes Energies,
IN2P3/CNRS, Universit\'es Paris VI \& VII, France 
\and Institut f\"ur Theoretische Physik, Lehrstuhl IV,
Ruhr-Universit\"at Bochum, Germany 
\and Institute of Particle and Nuclear Physics, Charles University,
Prague, Czech Republic
\and University of Namibia, Windhoek, Namibia 
\and European Associated Laboratory for Gamma-Ray Astronomy, jointly
supported by CNRS and MPG 
}

\date{Received ? / Accepted ?}

\abstract{We report on a possible association of the recently
discovered very high-energy $\gamma$-ray source HESS\,J1825--137 with
the pulsar wind nebula (commonly referred to as G\,18.0--0.7) of the
$2.1\times 10^{4}$ year old Vela-like pulsar
PSR\,B1823--13. HESS\,J1825--137 was detected with a significance of
8.1 $\sigma$ in the Galactic Plane survey conducted with the
H.E.S.S. instrument in 2004.
The centroid position of HESS\,J1825--137 is offset by 11\arcmin\,
south of the pulsar position. \emph{XMM-Newton} observations have
revealed X-ray synchrotron emission of an asymmetric pulsar wind
nebula extending to the south of the pulsar. We argue that the
observed morphology and TeV spectral index suggest that
HESS\,J1825--137 and G\,18.0--0.7 may be associated: the lifetime of
TeV emitting electrons is expected to be longer compared to the {\it
XMM-Newton} X-ray emitting electrons, resulting in electrons from
earlier epochs (when the spin-down power was larger) contributing to
the present TeV flux. These electrons are expected to be synchrotron
cooled, which explains the observed photon index of $\sim 2.4$, and
the longer lifetime of TeV emitting electrons naturally explains why
the TeV nebula is larger than the X-ray size.  Finally, supernova
remnant expansion into an inhomogeneous medium is expected to create
reverse shocks interacting at different times with the pulsar wind
nebula, resulting in the offset X-ray and TeV $\gamma$-ray morphology.

;

}
\titlerunning{The association of HESS\,J1825--137 with G\,18.0--0.7}
\maketitle

\section{Introduction}
\label{intro}
PSR\,B1823--13 (also known as PSR\,J1826--1334) is a 101 ms evolved
pulsar with a spin-down age of $T=2.1 \times 10^{4}$ years
\citep{Clifton} and in these properties very similar to the Vela
pulsar. It is located at a distance of $d=3.9\pm 0.4$
kpc~\citep{Cordes_Lazio} and {\it ROSAT} observations of this source
with limited photon statistics revealed a compact core, as well as an
extended diffuse nebula of size $\sim 5\arcmin$ south-west of the
pulsar~\citep{Finley}. High resolution {\it XMM-Newton} observations
of the pulsar region confirmed this asymmetric shape and size of the
diffuse nebula, which was hence given the name
G\,18.0--0.7~\citep{XMM}. For the compact core with extent
$R_{\mathrm{CN}}\sim 30\arcsec$ (CN: compact nebula) immediately
surrounding the pulsar, a photon index of
$\Gamma_{\mathrm{CN}}=1.6^{+0.1}_{-0.2}$ was measured with a
luminosity of $L_{\mathrm{CN}}\sim 9 d_4^2 \times 10^{32}$
erg\,s$^{-1}$ in the 0.5 to 10 keV range for a distance of $4d_4$
kpc. The corresponding pulsar wind shock radius is $R_s\leq 15\arcsec
= 0.3d_4$ pc. The compact core is embedded in a region of extended
diffuse emission which is clearly one-sided, revealing a structure
south of the pulsar, with an extension of $R_{\mathrm{EN}}\sim
5\arcmin$, (EN: extended nebula) whereas the $\sim 4\arcmin$ east-west
extension is symmetric around the north-south axis.  The spectrum of
this extended component is softer with a photon index of
$\Gamma_{\mathrm{EN}}\sim 2.3$, with a luminosity of
$L_{\mathrm{EN}}=3d_4^2\times 10^{33}$ erg\,s$^{-1}$ for the 0.5 to 10
keV interval. No associated supernova remnant (SNR) has been identified yet. 
 
At $\gamma$-ray energies, PSR\,B1823--13 was proposed to power the
close-by unidentified EGRET source 3EG\,J1826--1302~\citep{EGRET}. TeV
observations of this pulsar by the Whipple and HEGRA Collaborations
resulted in only upper limits~\citep{WhipplePSR1823, HEGRASCAN}, which
are unconstraining with respect to H.E.S.S.

The region around PSR\,B1823--13 was observed as part of the survey of
the Galactic plane with the H.E.S.S. instrument~\citep{HESSSCAN}. In
this survey, a source of very high-energy (VHE) $\gamma$-rays
(HESS\,J1825--137) 11$'$ south of the pulsar was discovered with a
significance of 8.1 $\sigma$. We note that the new VHE $\gamma$-ray
source is located within the 95\% positional confidence level of the
EGRET source 3EG\,J1826--1302 and could therefore be related to this
as of yet unidentified object. The High Energy Stereoscopic System
(H.E.S.S.) is an array of four imaging atmospheric Cherenkov
telescopes located in the Khomas Highland of Namibia~\citep{HESS}. It
is designed for the observation of astrophysical sources in the energy
range from 100 GeV to several tens of TeV. The system was completed in
December 2003 and has already provided a number of significant
detections of galactic $\gamma$-ray sources. These include the first
detection of spatially extended emission from a pulsar wind nebula
(PWN) in very high-energy $\gamma$-rays~\citep{HESSMSH}. Each
H.E.S.S. telescope has a mirror area of 107 m$^2$~\citep{HESSOptics}
and the system is run in a coincidence mode~\citep{HESSTrigger}
requiring at least two of the four telescopes to have triggered in
each event. The H.E.S.S. instrument has an energy threshold of
$\approx$ 100~GeV at zenith, an angular resolution of $\sim
0.1^{\circ}$ per event and a point source sensitivity of
$<2.0\times\,10^{-13}$ cm$^{-2}$s$^{-1}$ (1\% of the flux from the
Crab Nebula) for a $5\,\sigma$ detection in a 25 hour observation.

\section{H.E.S.S. Observations and Results}

The first H.E.S.S. observations of this region occurred as part of a
systematic survey of the inner Galaxy from May to July 2004 (with 4.2
hours of exposure within 2$^{\circ}$ of HESS\,J1825--137).  Evidence
for a VHE $\gamma$-ray signal in these data triggered re-observations
from August to September 2004 (5.1 hours).  The mean zenith angle of
the observations was 31$^{\circ}$ and the mean offset ($\psi$) of the
source from the pointing direction of the system was
$0.9^{\circ}$. The off-axis sensitivity of the system derived from
Monte-Carlo simulations has been confirmed via observations of the
Crab Nebula~\citep{HESSCrab}. The data set corresponds to a total
live-time of 8.4 hours after application of run quality selection
based on weather and hardware conditions.

\begin{figure}[h]
\centering
\includegraphics[width=8.5cm]{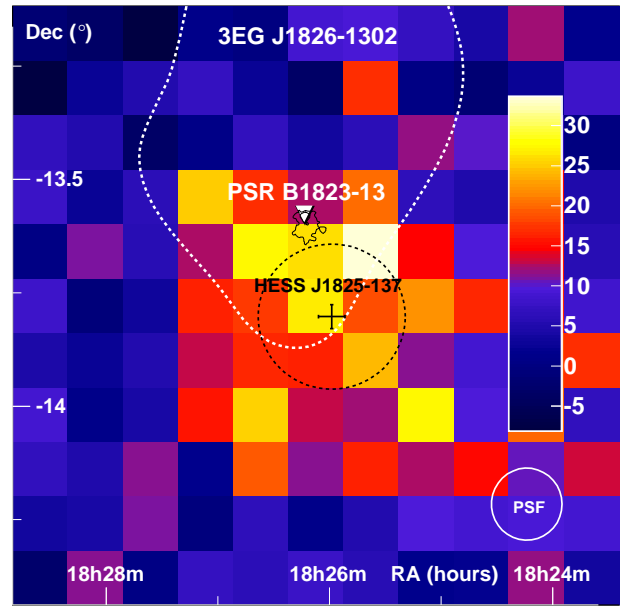}
\caption{ 
  Excess map of the region close to PSR\,B1823--13 (marked with a
  triangle) with uncorrelated bins. The best fit centroid of the
  $\gamma$-ray excess is shown with error bars. The black dotted
  circle shows the best fit emission region size
  ($\sigma_{\mathrm{source}}$) assuming a Gaussian brightness
  profile. The black contours denote the X-ray emission as detected by
  \emph{XMM-Newton}. The 95\% confidence region (dotted white line)
  for the position of the unidentified EGRET source 3EG\,J1826--1302
  is also shown. The system acceptance is uniform at the 20\% level in
  a $0.6^{\circ}$ radius circle around HESS\,J1825--137.}
\label{fig:skymap}
\end{figure}

\begin{figure}
\centering
\includegraphics[width=8.2cm]{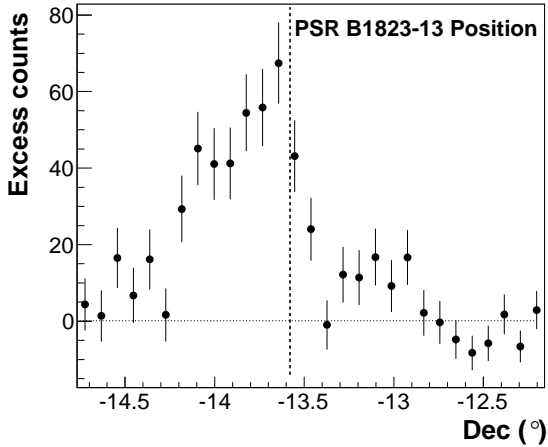}
\caption{Acceptance-corrected excess slice through the region
  surrounding HESS\,J1825--137 along the north-south direction of width
  0.4$^{\circ}$. The pulsar position is marked with a dotted line. The
  one-sided nature of the emission with a slow decline to the south of
  the pulsar is evident.}
\label{fig::ExcessSlice}
\end{figure}

The standard scheme for the reconstruction of events was applied to
the data (see \citealt{HESS2155} for details). Cuts on the scaled
width and length of images (optimized on $\gamma$-ray simulations and
off-source data) were used to suppress the hadronic background. While
in the standard scheme an image size cut of 80 photoelectrons (pe) was
used to ensure well reconstructed images, in the search for weak
sources an additional image size cut of 200 pe was applied to achieve
optimum sensitivity. This cut reduces the background by a factor of 7
at the expense of an increased analysis threshold of 420~GeV. A model
of the field of view acceptance, derived from off-source runs, is used
to estimate the background. Fig.~\ref{fig:skymap} shows an
uncorrelated excess count map of the $1.6^{\circ}\times1.6^{\circ}$
region around HESS\,J1825--137. A clear and extended excess is
observed to the south of the pulsar PSR\,B1823--13.  Assuming a
radially symmetric Gaussian brightness profile ($\rho \propto
\exp(-\theta^{2}/2\sigma_{\mathrm{source}}^2)$) an extension of
$\sigma_{\mathrm{source}}\,=\,9.6\arcmin \pm2.0\arcmin$ is derived.
The best fit position for the centre of the excess lies at a distance
of 11.2\arcmin\, from PSR\,B1823--13 at 18h26m3s$\pm$7s,
-13\degr 45.7\arcmin$\pm$1.7\arcmin. On the scale of the measured
displacement the systematic H.E.S.S. pointing uncertainties of
20\arcsec are negligible. Fig.~\ref{fig::ExcessSlice} shows an
acceptance corrected excess slice through the region surrounding
HESS\,J1825--137, in the north-south direction, of width
0.4$^{\circ}$. The position of PSR\,B1823--13 is marked with a dotted
line. It can be seen that the VHE emission extends asymmetrically to
the south of the pulsar. Using the best-fit position we derive a
statistical significance of 8.1~$\sigma$ after accounting for all
trials involved in the search for sources~\citep{HESSSCAN}.  This
significance is obtained counting events within a circle of radius
$\theta = 0.22^{\circ}$ ($\theta^2 = 0.05$ deg$^{2}$), a value chosen
a priori for the search for extended sources~\citep{HESSSCAN}. Using a
larger angular cut appropriate to contain the complete emission region
of HESS\,J1825--137 of $\theta = 0.4^{\circ}$ an excess of 370 $\pm$
20 and a statistical significance of 13.4 $\sigma$ are derived.

\begin{figure}
\centering
\includegraphics[width=0.5\textwidth]{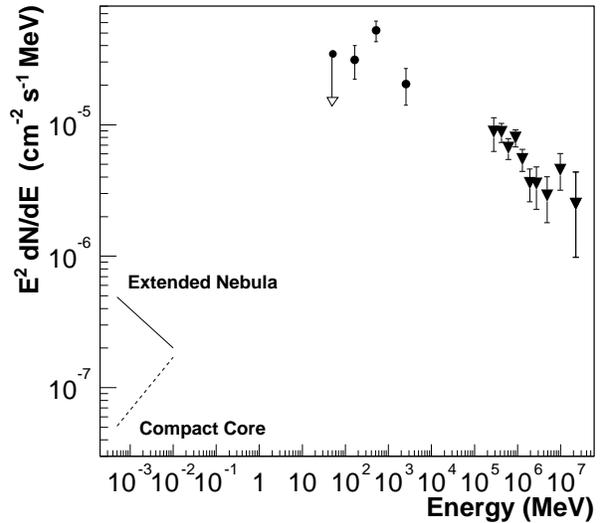}
\vspace{-0.2cm}
\caption{Spectral energy distribution of HESS\,J1825--137, assuming
  that the X-ray emission surrounding PSR\,B1823--13, the EGRET source
  3EG\,J1826--1302 and the new VHE $\gamma$-ray source are
  related. X-ray data, indicated by lines, are taken from~\citet{XMM}
  and are shown for the two different regions as described in the
  text. EGRET data (full circles) are taken from the third EGRET
  catalog~\citep{EGRETCat}. The triangles show the H.E.S.S. data
  from this work.}
\label{fig:spectrum}
\end{figure}

For spectral analysis, the looser image size cut of 80~pe and the wide
angular cut of $\theta < 0.4^{\circ}$ are applied to extend the source
spectrum to lower energies (resulting in a threshold of 230~GeV). To
reduce systematic errors, only runs with $\psi<1.5^{\circ}$ offset
from the on-region are used, resulting in a total livetime of 5.4
hours. The background is estimated from regions with equal offset
$\psi$ from the centre of the field of view, again to minimize
systematic errors. The derived spectral energy distribution is shown
in Fig.~\ref{fig:spectrum}. The H.E.S.S. spectrum can be fitted by a
power law in energy with photon index $2.40\pm0.09_{stat}\pm0.2_{sys}$
and a flux above 230~GeV of
$(3.4\pm0.2_{stat}\pm1.0_{sys})\,\times\,10^{-11}$ cm$^{-2}$s$^{-1}$
(corresponding to 12\% of the Crab flux above that energy). The fit
has a $\chi^2$/d.o.f. of 9.8/8. A spectral analysis using a 200~pe
image cut yields consistent results. We estimate the systematic error
on the absolute flux level to be 30\%. We find no evidence for
intra-night variability of the $\gamma$-ray flux. A fit of the
night-by-night $\gamma$-ray flux of the source to a constant value
yields a $\chi^2$/d.o.f. of 9.2/7.

\section{The association of HESS\,J1825--137 with G\,18.0--0.7}
\label{discussion}
HESS\,J1825--137 was discovered during an unbiased survey of the
galactic plane within the central $\pm 30\degr$ longitude sector.
Multiwavelength searches within a circle of radius
$\sigma_{\mathrm{source}}\sim 10\arcmin$ around the centre of gravity
revealed PSR\,B1823--13, at the edge of the source radius, as the only
plausible candidate counterpart. The discussion below investigates the
possibility of associating the pulsar and its nebula G18.0--0.7 with
HESS\,J1825--137.

The one-sided nature of this PWN as seen in TeV was already suggested by~\citet{XMM},
for the X-rays, based on the hydrodynamical simulations of~\citet{Blondin} for Vela X
and earlier studies referenced by Gaensler et al.: it was assumed that
the density of the medium surrounding the progenitor
star was inhomogeneous along the north-south direction, with the
density towards the northern direction significantly larger than
to the south. The reverse shock from the northern direction should then 
have crashed relatively early into the PWN, pushing the latter towards the south,
as observed. The apparent diameter $\sim 0.5\degr$ of the TeV source
indicates a relatively large PWN size, $R_{\mathrm{PWN}} = 17 d_4$ pc.
The unseen SNR shell in this scenario would have to be substantially
larger: in the simulations of \citet{Blondin}, as well as in a
sample of observed composite SNRs \citep{Swaluw_Wu}, the ratio
$R_{\mathrm{PWN}}/R_{\mathrm{SNR}}$ does not exceed $\sim 0.25$.
The large implied $R_{\mathrm{SNR}}$ would suggest that the
blast wave is expanding into a low-density medium to the south.
For instance, a remnant in the Sedov-Taylor phase expanding into the hot phase of the
interstellar medium, with density $\sim 0.003$ cm$^{-3}$,
would have $R_{\mathrm{SNR}} = 58$ pc at $T = 21.5$ kyears
(assuming an explosion energy of $10^{51}$ erg).
This is somewhat smaller than the value implied by the size
of the TeV source, but uncertainties in the distance estimate should
be kept in mind, as well as the fact that with a braking index
different from 3 (as discussed below), the true pulsar age might
be greater than the nominal spin-down time.
Within the scenario outlined above, a high initial spin-down luminosity,
or the fact that the SNR reverse shock on the southern side might
not yet have reached the PWN, could also yield a larger value for
$R_{\mathrm{PWN}}/R_{\mathrm{SNR}}$ \citep[][and references therein]{Bucciantini}.
Finally, the accelerated electrons might diffuse beyond the
boundary of the PWN, where they would still radiate by the
inverse Compton mechanism but emit little synchrotron radiation 
\citep{AhaAtoKif97}, making the TeV source extension larger than
otherwise be expected.

The synchrotron lifetime of VHE electrons in a field of strength
$B=10^{-5}B_{-5}$\,G, scattering cosmic microwave background (CMBR)
photons to energies $E_{\gamma}=10^{12}E_{\rm TeV}$\,eV (in the
Thomson limit) can be shown to be $\tau(E_\gamma)=4.8B_{-5}^{-2}E_{\rm
TeV}^{-1/2}$ kyears, whereas the corresponding lifetime of keV
emitting electrons is shorter:
$\tau(E_{\mathrm{X}})=1.2B_{-5}^{-3/2}E_{\rm keV}^{-1/2}$ kyears,
where $E_{\rm keV}$ is the synchrotron photon energy in units of keV.
Gaensler et al.\ have argued that the field strength in the extended
X-ray nebula is $\sim$ 10 $\mu$G, so that we assume $B_{-5}=1$ for the
X-ray and TeV emitting zones.  The adiabatic loss and escape time
scales from the compact nebula are a few years; this is much shorter
than the corresponding synchrotron lifetime of X-ray emitting
electrons, given typical values of Vela-like compact nebular field
strengths.  Synchrotron losses in the compact nebula therefore do not
modify the electron spectral index of $\sim 2.2$ in this region, which
is responsible for the observed compact nebular photon index of
$\Gamma_{\mathrm{CN}}\sim 1.6$ in the 0.5 to 10 keV range.  The
shocked pulsar wind particles in the offset nebula will then propagate
southwards until the propagation time is equal to the synchrotron loss
time, which will result in a steepening in the X-ray photon index to
the observed value of $\Gamma_{\mathrm{EN}}\sim 2.3$. If $V_{\gamma}$
and $V_{\rm X}$ are the average wind convection speeds corresponding
to the respective $\gamma$-ray and X-ray emitting zones, the ratio
between the predicted TeV and X-ray sizes should then be
$R_{\gamma}/R_{\rm X}=4(V_{\gamma}/V_{\rm X})B_{-5}^{-1/2}
(\overline{E}_{\rm keV}/\overline{E}_{\rm TeV})^{1/2}$, which can
easily predict a TeV nebula which is $\sim 6$ times larger than the
X-ray nebula, given the respective experimental mean X-ray and
$\gamma$-ray photon energies of $\overline{E}_{\rm keV}\sim 2$ (after
absorption) and $\overline{E}_{\rm TeV}\sim 0.9$.  This ratio (of six)
also assumes that the expansion velocity of the PWN does not change
significantly between the extended X-ray and TeV nebulae
(i.e. $V_{\gamma}\sim V_{\rm X}$). It should however be noted that
this is a naive assumption and that a detailed study of the evolution
of the velocity and associated magnetic field distributions should be
made.

Pulsars spin down with a braking law given by
$\dot{\Omega}=-K\Omega^n$, where $\Omega$ is the spin angular
frequency and $K$ is a constant, depending on the surface magnetic
field strength and neutron star equation of state. Assuming typical
pulsar braking indices $n\sim 2.5$ to 3, we can integrate over the
spindown power $\dot{E}$ (assuming a conversion efficiency of $\sim 50\%$
to electrons) to an epoch
$\tau(E_\gamma)$ into the past when $\dot{E}$ was larger, to give the
total electron spectrum contributing to the H.E.S.S. spectral band
(from 0.23 TeV to $ > 10$ TeV). Since we integrate over the past
$\tau_{\mathrm{max}} \sim 9 B_{-5}^{-2}$ kyears (corresponding to a
minimum energy of 0.23 TeV), which represents $\sim 50\%$ of the
pulsar lifetime, uncertainties in $n$ should still be relatively
unimportant.  The total observed TeV spectrum is then the result of a
summation of successive inverse Compton spectra arising from past
injected electron spectra, with the spectral break energy shifting
down with time $T$ into the past as $T^{-2}$.  The result is a cooled
TeV spectrum, for which the TeV $\gamma$-ray photon index
$\Gamma_{\mathrm{TeV}}$ should be larger than
$\Gamma_{\mathrm{CN}}+0.5\sim 2.1$ due to second-order Klein-Nishina
effects on the production spectrum of IC radiation on the CMBR.  A
detailed discussion of this is beyond the scope of this paper, but the
important point is that the observed TeV photon index in the range 2.2
to 2.6 is consistent with this interpretation. Furthermore, because
the TeV emission is the result of earlier epochs of pulsar injection,
the ratio of $\gamma$-ray luminosity to present spin-down power
would over-estimate the true conversion efficiency. Compare the X-ray and
$\gamma$-ray energy fluxes in Fig.~\ref{fig:spectrum}: the TeV energy
flux is larger than the X-ray energy fluxes, with the latter resulting
from more freshly injected electrons.

EGRET did not detect pulsed emission above 100 MeV from this
pulsar~\citep{Nel} and phase-resolved spectroscopy is required to set
upper limits on the pulsed component associated with PSR\,B1823--13,
to see if this component is significantly lower than the EGRET steady
excess shown in Fig.~\ref{fig:spectrum}. If 3EG\,J1826--1302 is also
associated with G\,18.0--0.7, the GeV emitting electrons would
represent the earliest epochs of pulsar injection. This emission
should also be one-sided within the framework discussed above and
future GLAST observations may be able to provide better constraints on
the GeV morphology.

\section*{Acknowledgements}

The support of the Namibian authorities and of the University of Namibia
in facilitating the construction and operation of H.E.S.S. is gratefully
acknowledged, as is the support by the German Ministry for Education and
Research (BMBF), the Max Planck Society, the French Ministry for Research,
the CNRS-IN2P3 and the Astroparticle Interdisciplinary Programme of the
CNRS, the U.K. Particle Physics and Astronomy Research Council (PPARC),
the IPNP of the Charles University, the South African Department of
Science and Technology and National Research Foundation, and by the
University of Namibia. We appreciate the excellent work of the technical
support staff in Berlin, Durham, Hamburg, Heidelberg, Palaiseau, Paris,
Saclay, and in Namibia in the construction and operation of the
equipment. 


\begin{thebibliography}{}
\bibitem[Aharonian et al.(1997)]{AhaAtoKif97}Aharonian, F.A.,
Atoyan, A., \& Kifune, T. 1997, MNRAS 291, 162
\bibitem[Aharonian et al.(2002)]{HEGRASCAN}Aharonian, F.A. et al.,
({\it HEGRA Collaboration}) 2002, A\&A, 395, 803
\bibitem[Aharonian et al.(2005a)]{HESSSCAN}Aharonian, F.A. et al.,
({\it H.E.S.S. Collaboration}) 2005a, Science, 307, 1938
\bibitem[Aharonian et al.(2005b)]{HESSMSH}Aharonian, F.A. et al., ({\it
H.E.S.S. Collaboration}) 2005b, A\&A, 435, L17
\bibitem[Aharonian et al.(2005c)]{HESSCrab}Aharonian, F.A. et al.,
({\it H.E.S.S. Collaboration}) 2005c, to be submitted to A\&A
\bibitem[Aharonian et al.(2005d)]{HESS2155}Aharonian, F.A. et al.,
({\it H.E.S.S. Collaboration}) 2005d, A\&A, 430, 865
\bibitem[Bernl\"ohr et al.(2003)]{HESSOptics}Bernl\"ohr, K. et
al. 2003, Astropart. Phys., 20, 111
\bibitem[Blondin et al.(2001)]{Blondin}Blondin, J.M., Chevalier, R.A.,
\& Frierson, D.M. 2001, ApJ, 563, 806
\bibitem[Bucciantini et al.(2004)]{Bucciantini}Bucciantini, N., Bandiera, R.,
Blondin, J.~M., Amato, E., \& Del Zanna, L. 2004, A\&A, 422, 609
\bibitem[Clifton et al.(1992)]{Clifton}Clifton, T.~R., Lyne, A.~G., Jones, A.~W., McKenna, J., \& 
Ashworth, M. 1992, MNRAS, 254, 177
\bibitem[Cordes \& Lazio(2002)]{Cordes_Lazio}Cordes, J.~M., \& Lazio,
T.~J.~W. 2002, preprint (astro-ph/0207156)
\bibitem[Finley et al.(1998)]{Finley}Finley, J.~P., Srinivasan, R., \&
 Park, S. 1996, 466, 938
\bibitem[Funk et al.(2004)]{HESSTrigger}Funk, S. et al., 2004,
Astropart. Phys., 22, 285
\bibitem[Gaensler et al.(2003)]{XMM}Gaensler, B.~M., Schulz, N.~S.,
Kaspi, V.~M.  Pivovaroff, M.~J., \& Becker, W.~E. 2003, ApJ, 588, 441
\bibitem[Hall et al.(2003)]{WhipplePSR1823}Hall, T.~A., et al. 2003,
Proc. 28th ICRC, Tsukuba, Univ. Academy Press, Tokyo, p. 2497
\bibitem[Hartman et al.(1999)]{EGRETCat} Hartman, R.~C., et al., 1999,
  Astrophys. J. Suppl. Ser. 123, 79
\bibitem[Hinton(2004)]{HESS}Hinton, J.A. ({\it
H.E.S.S. Collaboration}) 2004, New Astron. Rev., 48, 331
\bibitem[Nel et al.(1996)]{Nel}Nel, H.I. et al. 1996, ApJ, 465, 898
\bibitem[Nolan et al.(2003)]{EGRET}Nolan, P.~L, Tompkins, W.~F.,
Grenier, I.~A., \& Michelson, P.~F. 2003, ApJ., 597, 615
\bibitem[van der Swaluw \& Wu (2001)]{Swaluw_Wu}van der Swaluw, E., \& Wu, Y. 2001, ApJ, 555, L49
\end{thebibliography}
\end{document}